# Exploration of Artificial Intelligence-oriented Power System Dynamic Simulators

Tannan Xiao, Ying Chen, Jianquan Wang, Shaowei Huang, Weilin Tong, Tirui He

*Abstract*—With the rapid development of artificial intelligence (AI), it is foreseeable that the accuracy and efficiency of dynamic analysis for future power system will be greatly improved by the integration of dynamic simulators and AI. To explore the interaction mechanism of power system dynamic simulations and AI, a general design of an AI-oriented power system dynamic simulator is proposed, which consists of a high-performance simulator with neural network supportability and flexible external and internal application programming interfaces (APIs). With the support of APIs, simulation-assisted AI and AI-assisted simulation form a comprehensive interaction mechanism between power system dynamic simulations and AI. A prototype of this design is implemented and made public based on a highly efficient electromechanical simulator. Tests of this prototype are carried out under four scenarios including sample generation, AI-based stability prediction, data-driven dynamic component modeling, and AI-aided stability control, which prove the validity, flexibility, and efficiency of the design and implementation of the AI-oriented power system dynamic simulator.

*Index Terms*—Power system dynamic simulator, artificial intelligence, application programming interface, parallel computing.

## I. Introduction

### A. Motivations

POWER system dynamic simulation is still the most reliable and widely used approach for power system stability analysis [1]. Electric power companies and developers from all over the world have developed many dynamic simulators including electromechanical simulators such as PSASP [2] and PSD-BPA [3] by the China Electric Power Research Institute (CEPRI), PSS/E [3] by Siemens, DSATools [4] by Powertech, DIgSILENT PowerFactory [5] by DIgSILENT GmbH, Eurostag [6] by Tractebel, PYPOWER-Dynamics [7] by Susanto, STEPS [8] by Shandong University, etc., electromagnetic simulators such as PSCAD/EMTDC [9] by Manitoba, CloudPSS [10] by Tsinghua University, etc., and real-time simulators such as RTDS [11]

Manuscript received: February 23, 2022; accepted: XX XX, XXXX. Date of CrossCheck: XX XX, XXXX. Date of online publication: XX XX, XXXX.

This work was supported in part by the National Natural Science Foundation of China under Grants 52107104, 51877115, and 51861135312, and in part by China Postdoctoral Science Foundation under Grant 2021M691724.

Tannan Xiao, Ying Chen (corresponding author), Shaowei Huang, Tirui He are with the Department of Electrical Engineering, Tsinghua University, Beijing, 100084, China (Email: eexiaoxh@gmail.com, chen_ying@tsinghua.edu.cn, huangsw@tsinghua.edu.cn, hetirui@qq.com).

Jianquan Wang is with the College of Electrical Engineering, Zhejiang University, Hangzhou, 310027, China (Email: wangjq@zju.edu.cn).

Weilin Tong is with Wuxi Power Supply Company of State Grid, Wuxi, China (Email: tongwl1994@qq.com).

by Manitoba, HYPERSIM [12] by OPAL-RT, ADPSS [13] by CEPRI, etc. The commercial simulators are well tested in the practical power systems, which means they support a lot of functions and are very reliable. However, the commercial simulators are usually designed and implemented years ago, which means their architecture might be old and the application programming interfaces (APIs) may be stiff or even be unavailable. On the other hand, the free and open-source simulators are commonly not as functionally mature as the commercial ones but are much more flexible. The source code can be directly modified, so APIs can be developed as needed.

Research on artificial intelligence (AI) has achieved a growth spurt in the past few years. AI algorithms such as graph neural networks (GNN), reinforcement learning (RL), etc., have been applied to a variety of power system studies such as measurement enhancement [14], dynamic component modeling [15], parameter inference [16], optimization and control [17], stability assessment [18], etc. AI models can learn and approximate any functions with enough samples [19]. AI technology will be more and more important in the research field of power systems, especially with the rapid development of renewable generation and power electronics [20]. The safe and efficient operation of power systems is facing great challenges, e.g., we may need to model new devices to analyze stability although some devices' operating mechanisms are still under research, dimensionality reduction is needed to scale down the complexity, the significantly increased uncertainty of power systems requires fast and flexible stability analysis and control, etc. AI-assisted power system analysis and control might be a solution to these challenges [21], or at least a mitigation measure.

Currently, although some commercial software such as DIgSILENT PowerFactory has supported AI applications to a certain extent, the relationship between power system dynamic simulation and AI is still relatively fragmented. The simulator usually works only as a data generator and provides limited prior knowledge, whereas the trained AI model usually works independently as a black-box model with poor interpretability and cannot be easily integrated into simulators. In one word, the simulator is not AI-friendly enough. In [22], the idea of a learning simulation engine that combines AI and simulation is proposed. Simulation-assisted AI and AI-assisted simulation mutually support each other. With the support of AI, the simulator can evolve autonomously and become more accurate and efficient, which is also crucial to the realization of power system digital twins [23]. In [24], a similar and more comprehensive concept of simulation intelligence is proposed. This can be the future of power system dynamic simulators.

## B. Contributions

Based on the learning simulation engine proposed in [22] and the simulation intelligence discussed in [24], we explore the cooperation of AI and power system dynamic simulations in this paper. The contributions are as follows.

1) A general design of an AI-oriented power system dynamic simulator is proposed, which consists of a high-performance simulator with neural network supportability and flexible external and internal APIs. External API-based Simulation-assisted AI and internal API-based AI-assisted simulation form a comprehensive interaction mechanism between power system dynamic simulations and AI.

2) A prototype of the proposed design is implemented based on a self-developed power system electromechanical simulator. The efficiency of the simulator is improved by traditional approaches including sparsity techniques, parallel computing, and memory allocation optimization. External and internal APIs are developed with Python and the source code is made public on GitHub[1].

3) Four typical cases of utilizing the developed simulator, i.e., sample generation, GNN-based stability prediction, data-driven dynamic component modeling, and RL-based stability control, are illustrated to prove the validity, flexibility, and efficiency of the proposed design and implementation. All four cases are supported by at least one paper or the source code we published on GitHub.

## C. Paper Organization

The remainder of the papers is as follows. Section II introduces the design of the AI-oriented power system dynamic simulator and discusses the interaction of dynamic simulations and AI. The implementation details of a prototype simulator based on the proposed design are illustrated in Section III. In section IV, the typical examples of the implemented simulator are explained and tested. Conclusions are drawn in Section V.

## II. DESIGN OF AI-ORIENTED DYNAMIC SIMULATOR

In Fig. 1, the overall architecture design of the AI-oriented simulator is demonstrated. The idea is intuitive. In order to support the interactions between the simulator and AI, a reasonable choice is to develop AI-friendly APIs to bridge the two. The simulator, APIs, AI models, and power system operator form a bionic interaction mechanism similar to the musculoskeletal system, nerves, spinal cord, and brain. Via the APIs, the simulator can provide massive data and prior knowledge for AI models, whereas AI models mine the data, discover the hidden patterns, and return well-trained models and posterior knowledge. Meanwhile, AI models can provide predictions and suggestions to power system operators based on their demands. Therefore, a closed-loop interaction mechanism is established. With the support of this interaction mechanism, the traditional physics-based simulator and the data-driven AI models can cooperate to achieve the task of data enhancement [25], awareness enhancement [26], analysis enhancement [27], decision-making enhancement [28], etc., and may finally lead to the creation of a power system digital twin.

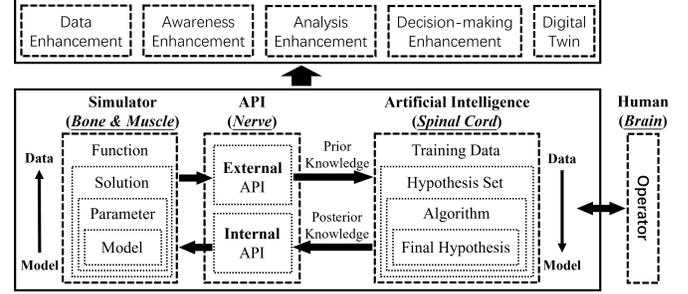

Fig. 1. The overall architecture of the AI-oriented simulator.

In this section, the designs of the simulator and APIs, as well as the interactions of the simulator and AI models are explained in detail.

### A. Simulator

Power system dynamic simulators can be used to generate massive scenarios and simulation results, i.e., physics-based data. The general structure of a power system simulator is displayed in Fig. 1, which consists of four parts, namely, model, parameter, solution, and function. Firstly, a dynamic model is needed for each power system component. It can be a physics-based model, a data-driven model, or a physics-data-integrated model. A model conversion function for different models of different simulators is preferred. Secondly, parameters of the selected model need to be measured or estimated, i.e., model calibration [29] is needed. Thirdly, all the models with parameters are formulated together in a group of high-dimensional equations. Power flow can be solved with the Newton method to obtain the operation state. Power system dynamics are formulated with ordinary differential equations (ODEs) in the electromagnetic simulation and differential-algebraic equations (DAEs) in the electromechanical simulation. They can both be solved with a numerical integration method and a linear solver [30]. Finally, simulation-based functions are realized based on the solutions of power flow and power system dynamics.

Here are two required features of the simulator to support the integrations of simulations and AI models, i.e., rapid simulation speed and neural network supportability.

*1) Rapid Simulation Speed*

Simulation speed is essentially the basis of AI-assisted power system analysis and control. The training of AI models requires massive data. The data generation is very time-consuming. The simulation speed is a bottleneck in successfully utilizing AI algorithms and training a model with sufficient performance. Therefore, the simulator must be well optimized and highly efficient. Algorithm-level and task-level parallelism, which is solution-level and function-level in Fig. 1, is required to fulfill the efficiency needs in different situations.

*2) Neural Network Supportability*

Another requirement of the simulator is neural network supportability, i.e., being able to load the structure and parameters of neural networks and perform at least forward propagation of neural networks. The simulator should be able to integrate AI models into any part of the simulator so that the efficiency and accuracy of the simulator can be improved by the cooperation of traditional algorithms and AI.

---
[1] https://github.com/xxh0523/Py_PSOPS.

### B. Application Programming Interface (API)

Flexible APIs are crucial to the cooperation of power system simulations and AI models. The APIs of an AI-oriented power system simulator are divided into external APIs and internal APIs. External APIs are used to get data and invoke functions of the simulator, whereas internal APIs are used to modify data and control the simulation process. Corresponding to the four parts of the simulator shown in Fig. 1, the external and internal APIs can further be divided into four categories, namely, model APIs, parameter APIs, solution APIs, and function APIs.

External model APIs and parameter APIs are used to output structures and parameters of different models. The model expression should be easy to understand and modify, e.g., diagrams, JSON files, etc. External solution APIs are used to output parameters related to solution methods and intermediate results during the solution process, e.g., the node ordering method, the admittance matrix, the iteration number of the power flow solution, the integration step, etc. External function APIs are used to invoke functions e.g., power flow solutions, short-circuit calculations, transient simulations, etc., as well as output required simulation results, e.g., the maximum rotor angle difference, nodal voltages, currents of transmission lines, etc.

Correspondingly, the internal APIs are used to alter model types, modify parameters, change solution methods, and adjust boundary conditions of functions. Through internal APIs, AI models can be used to model dynamic components, estimate parameters, accelerate solutions, and surrogate functions.

Overall, compared with the APIs of existing commercial simulation tools, the APIs of an AI-oriented power system simulator make it easy to develop AI models for power system applications and realize the cooperation of traditional algorithms and AI algorithms. More detailed information on power system components and simulations can be provided to AI models via the external APIs and AI models can be deeply integrated into power system simulations via the internal APIs. Other suggested features of the APIs are as follows.

*1) No Impact on Simulation Efficiency*

The implementation of APIs must not affect the efficiency of the simulator. As mentioned before, the simulator focuses on efficiency. The source code is usually written with efficient programming languages such as C++, Java, FORTRAN, etc., i.e., the implementation is highly organized and optimized. It should not be disturbed by the APIs. Therefore, a suggested way is to rewrap the needed internal functions as external functions. These external functions can be invoked by other programming languages.

*2) Efficient Memory Exchange*

Data are frequently exchanged between the simulator and AI models. Taking neural network-based stability prediction as an example, massive samples of tens or even hundreds of gigabytes are generated. The data exchange better happens in RAM instead of on hard drives. If there is not enough RAM, the data could be cut into several pieces and transferred sequentially, or the data could be exchanged using a database.

*3) Interpreted Language-Written*

It is recommended to develop the APIs of an AI-oriented power system simulator with interpreted languages such as Python. These programming languages are easy to learn and use, and have very mature developer communities. There are a tremendous number of Python-written open-source AI projects on GitHub. With Python APIs, the simulator can be easily modified for AI applications and cooperate with AI models.

### C. Interactions of Simulations and AI

In contrast to the simulator, AI produces data-driven models based on existing data, as illustrated in Fig. 1. Firstly, a training dataset is needed. The quality and representativeness of the samples will seriously affect the performance of AI models. Secondly, a hypothesis set is established, i.e., a learning framework is selected based on the task and the training data. Thirdly, optimization algorithms are utilized to train the model. At last, the final hypothesis, i.e., an AI model, is obtained.

The interactions between the simulator and AI models are further illustrated in Fig. 2. The right part demonstrates simulation-assisted AI and the left part denotes AI-assisted simulation.

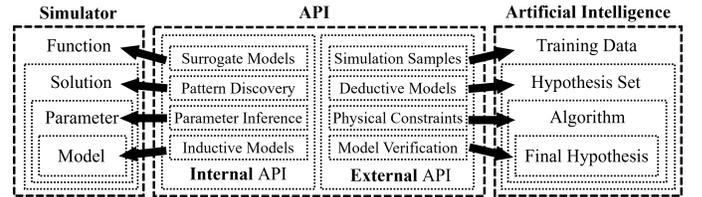

Fig. 2. Interactions between simulation and AI.

*1) Simulation-Assisted AI*

Simulation-assisted AI is realized based on the external APIs. Firstly, simulations can be used to generate training data. The actual training dataset also needs sample selection or augmentation, e.g., stability prediction needs simulation results with balanced stability labels, etc. Sampling methods are very important since the data quality determines the performance of AI models [31]. Secondly, physics-based models can be used as a strong prior knowledge for AI model design. For example, the power network topology can be used to design GNN [18]. Thirdly, physical laws such as conservation laws can be used as constraints in optimization algorithms to limit the feasible region, improve the interpretability of AI models, and speed up the training process [32], [33]. Finally, the simulator can be used as the benchmark for the performance verification of AI models.

*2) AI-Assisted Simulation*

AI-assisted simulation is realized based on the internal APIs. Firstly, data-driven AI models can be used for dynamic component modeling [15], [34]. Although the model may suffer from the problem of interpretability, the measurement-based AI model can also be accurate and adaptive. Via the internal model APIs, physics-based models and data-driven models can be integrated seamlessly and be simulated simultaneously [35]. Secondly, AI models can be used for power system model calibration [36]. Power system dynamic modeling and parameter estimation are facing increasing challenges because of the rapid development of renewable generation and power electronics. Data-driven power system dynamic modeling could be a potential solution. Thirdly, AI models can be used to discover and formulate the hidden patterns in the solution procedures and improve solution efficiency, e.g., convergence prediction of power network equations, variable integration step prediction, switch state prediction of power electronic devices [37], etc. Finally, AI models can be used as surrogate models for power system analysis [38] and control [39]. Power system computation can be very time-consuming. Using a surrogate model as an approximation of the actual computation can significantly increase the efficiency of analysis and decision-making.

## III. IMPLEMENTATION OF AN AI-ORIENTED SIMULATOR

In Fig. 3, a prototype of the AI-oriented simulator explained in the former section is implemented based on a high-performance electromechanical simulator called Power System Optimal Parameter Selection (PSOPS). After developing some external functions to support Python APIs, the simulator is compiled as a dynamic link library PSOPS.dll in Windows and a shared object file PSOPS.so in Linux. The Python APIs of the prototype are developed using the ctypes library [40]. The PSOPS.dll, PSOPS.so, and the open-source Python APIs, can be found in the repository called Py_PSOPS on GitHub.

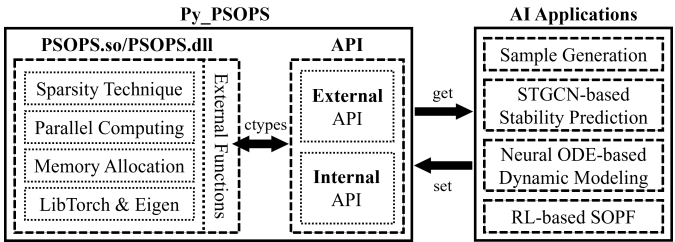

Fig. 3. Implementation and tests of Py_PSOPS.

In this section, the implementation details of the PSOPS simulator and the APIs are illustrated.

### A. Implementation of PSOPS

PSOPS can perform AC power flow considering PV-PQ switching and electromechanical transient stability simulations. It is developed using C++ based on previous studies [41], [42], [43], and [44]. Power system dynamics are modeled with a group of high-dimensional nonlinear differential-algebraic equations. The alternating approach proposed in [45] is adopted in PSOPS due to its simplicity, reliability, and robustness [30].

*1) Traditional Techniques for Efficiency Improvement*

In PSOPS, traditional algorithms including improved sparsity techniques, improved bordered block diagonal form (BBDF) method, and memory allocation techniques are utilized to accelerate transient simulations.

As for the sparsity techniques, the Approximate Minimum Degree Minimum Number of Source Predecessors (AMD-MNSP) algorithm and the multi-path sparse vector method are utilized to enhance the efficiency of the sparse vector method while maintaining the sparsity of the factorized matrix [41].

In terms of parallel computing, a fully parallel BBDF method [42], a fully parallel nested BBDF method [43], and an efficient computing task allocation scheme based on subnet-core mapping and mixed programming of MPI and OpenMP [44] are adopted to improve the BBDF method at the algorithmic and implementational levels.

As for memory allocation, a node ordering-based memory allocation technique is utilized, which reorders components such as transmission lines, transformers, generators, and loads based on the node ordering and saves the admittance matrix, the equation coefficient matrix, and the independent vector in a contiguous memory block to increase the cache hit rate when solving power network equations.

In TABLE I, the basic information and average time consumption of 10-second transient simulations after utilizing the above techniques are displayed. The 2383wp system is a widely used test system in MATPOWER. Sys13490 and Sys24886 are two practical power systems. All the dynamic components are modeled in detail. The test HPC platform is Sugon I950r-G installed with 8 Intel Xeon E7-8837 2.67 GHz processors. Each processor is integrated with 8 CPU cores, i.e., the total number of CPU cores is 64. As can be seen, simulations are significantly accelerated.

TABLE I
BASIC INFORMATION AND AVERAGE TIME CONSUMPTION OF 10-SECOND SIMULATIONS OF THREE TEST SYSTEMS

| Test Systems | Number of components | | | | Time Consumption (seconds) | |
|---|---|---|---|---|---|---|
| | Bus | Branch | Generator | Load | Serial | Parallel |
| 2383wp | 2383 | 2892 | 327 | 1822 | 2.655 | 0.365 |
| Sys13490 | 13490 | 22544 | 1797 | 3550 | 9.911 | 0.587 |
| Sys24886 | 24886 | 39512 | 1919 | 5646 | 13.525 | 0.639 |

*2) C++ Libraries for Neural Network Supportability*

Meanwhile, neural network supportability is realized using two open-source C++ libraries, namely, the Eigen library [46] and the LibTorch library [47]. The difference between the two libraries is the granularity of neural network implementation. As for the Eigen library, neural networks are built in the source code of the PSOPS simulator using the vector class and the matrix class. The structure of neural networks is loaded by reading a JSON file and the parameters are loaded by reading binary files saved by PyTorch [48]. On the other hand, the neural modules saved by PyTorch, i.e., both the structures and the parameters, are directly loaded for the LibTorch library. The Eigen library is computationally more efficient but structural changes of neural networks require modification of the source code of PSOPS, whereas the LibTorch library is simpler to use but the computation speed is lower than the Eigen library. Therefore, the LibTorch library is recommended for model evaluation and the Eigen library is recommended for model deployment. By changing the version of LibTorch, Py_PSOPS can adapt to different versions of PyTorch.

As mentioned before, neural network supportability makes it possible to integrate AI with traditional simulations. Currently, the integration and simultaneous simulations of AI-based power system dynamic models and physics-based models have been realized in PSOPS. Other applications of AI-assisted simulations are still under development.

### B. Implementation of APIs

As shown in Fig. 3, the PSOPS simulator is compiled as a dynamic link library file PSOPS.dll in Windows and a shared object file PSOPS.so in Linux. Only the external functions, which are realized by rewrapping the models and functions of PSOPS, can be accessed. The external functions of PSOPS and the Python APIs are connected using the ctypes library.

The Python APIs load the external functions from PSOPS.dll and PSOPS.so and reorganize the data into a NumPy [49] style. The source code is organized in a component-based manner, which means the functions of the same kind of component are put together. In the Python APIs, computational functions such as power flow solution and transient simulation have names that begin with "cal". Functions with names that start with "get" and "set" denote the functions of external APIs and internal APIs respectively. The Python APIs can be extended easily to fulfill the needs in different situations. A more well-rounded API will be a future working direction. The details of external and internal model APIs, parameter APIs, solution APIs, and function APIs are as follows.

*1) Model API*

As mentioned before, when using the Eigen library, the structure of neural networks can be established in the simulator by modifying the basic computation data file and reading a JSON file containing the names and structure of layers of the neural network. When using the LibTorch library, the whole neural model can be established by directly loading the modules saved by PyTorch.

*2) Parameter API*

Components' parameters such as the name, the total number, the constraints, the default settings, dynamic model parameters, connectivity, etc., can be got or set. However, currently, the parameters of neural networks are directly loaded by the simulator via modifying the basic computation data file and reading a binary file.

*3) Solution API*

The intermediate results during simulation processes can be reached. The simulated power system can be set to state at any integration step. Basic data of the solutions such as the iteration number, the simulation time, the integration step, faults, disturbances, etc., can be accessed. More importantly, the network topology accessibility is realized. Network topology data such as the admittance matrix, the impedance matrix, the number of fill-ins, and the factorized lower and upper triangular matrices can be obtained. The connectivity of each component, i.e., whether the component is connected to the power network, can be changed and network connectivity check is supported, i.e., asynchronous subsystems can be identified. Other settings such as power flow solution methods, integration methods, node ordering algorithms, and sparse vector methods can be modified by changing the basic computation data file.

*4) Function API*

The function API supports calling power flow solutions and transient stability simulations and gets simulation results including rotor angles, rotation speed, inner electric potential, active and reactive power, regulators' outputs, nodal voltages, etc. Meanwhile, task-level parallelism is realized using the ray library [50] of Python.

## IV. CASE STUDIES

In this section, four typical cases of utilizing the prototype, namely, sample generation, spatiotemporal graph convolutional networks (STGCN)-based stability prediction, neural ODE-based [51] power system dynamic component modeling, and RL-based stability-constrained optimal power flow (SOPF), are demonstrated to show the simulator-AI interactions based on Py_PSOPS, as shown in Fig. 3. Sample generation is one of the most basic applications of Py_PSOPS. The STGCN-based stability prediction, the neural ODE-based dynamic modeling, and the RL-based SOPF are typical examples of simulation-assisted AI, AI-assisted simulation, and mutual assistance between simulation and AI, respectively. More applications can be developed using Py_PSOPS. All four cases are supported by at least one paper or open-source code we developed on GitHub. The success of these tasks proves the validity, flexibility, and efficiency of the design and implementation.

The test system is the IEEE-39 system. The high-performance server used for testing consists of an NVIDIA P100 GPU, 250 gigabytes of RAM, and two Intel Xeon Gold 5118 processors, which contain 24 CPU cores in total and hyperthreading is enabled, i.e., there are up to 48 threads available.

### A. Sample Generation

*1) Stepwise Power Flow Sampling Scheme*

Sample generation can be used for any AI application. It is supported by the rapid simulation speed of PSOPS. As for power flow sampling, simple random sampling, grid sampling, and a stepwise sampling scheme are implemented. The pseudo-code of the stepwise scheme is shown as follows.

---
**Pseudo-code 1: Stepwise Power Flow Sampling Scheme**

**input** the total number of required samples $N$
**set** $n=1$
**get** the limits of $\mathbf{P}_D$, $\mathbf{Q}_D$, $\mathbf{P}_G$, and $\mathbf{V}_G$ using **external parameter APIs**
**while** $n<N$:
    **set** the connectivity states of all the transmission lines to **true** using **internal solution APIs**
    **set** the connectivity state of one or two randomly chosen transmission lines to **false** using **internal solution APIs**
    **set** $\mathbf{P}_D$ and $\mathbf{Q}_D$ randomly within limits using **internal parameter APIs**
    **calculate** $\text{sum}(\mathbf{P}_D)$
    **do**
        **set** $\mathbf{P}_G$ randomly within limits using **internal parameter APIs**
    **until** $\text{sum}(\mathbf{P}_G)+\text{sum}(\underline{\mathbf{P}}_{Slack})<\text{sum}(\mathbf{P}_D)<\text{sum}(\mathbf{P}_G)+\text{sum}(\overline{\mathbf{P}}_{Slack})$
    **set** $\mathbf{V}_G$ randomly within limits using **internal parameter APIs**
    **power flow** calculation using **external function APIs**
    **get** power flow convergence using **external solution APIs**
    **if** power flow calculation converges:
        **save** ($\mathbf{P}_D$, $\mathbf{P}_G$, $\mathbf{V}_G$)
        **set** $n \leftarrow n+1$
    **end if**
**end while**

---

where $\mathbf{P}_D$ and $\mathbf{Q}_D$ are the active power vector and reactive power vector of loads, respectively, $\text{sum}(\bullet)$ denotes the sum of elements in the vector, $\mathbf{P}_G$ is the active power vector of generators, $\overline{\mathbf{P}}_{Slack}$ and $\underline{\mathbf{P}}_{Slack}$ are the upper limit vector and the lower limit vector of slack generators, $\mathbf{V}_G$ is the nodal voltage vector of generators other than slack generators. The specific APIs used in each step are highlighted in bold.

After power flow sampling, contingencies are sampled by randomly choosing a component, a fault type, a fault location,

and a fault clearing time. Transient simulations of these contingencies are carried out to generate simulation samples.

*2) Test Results*

The source code of the proposed sampling scheme can be found on GitHub in the Py_PSOPS repository. On the test server, over 1.29 million power flow samples and over 50 million simulation samples of the IEEE-39 system are generated using 40 threads within 9 hours. This sample dataset is used to support the research on STGCN and neural ODE.

*B. STGCN-based Stability Prediction*

This is a typical example of simulation-assisted AI. The simulator provides training data as well as prior knowledge to support AI model design.

*1) STGCN*

An STGCN-based stability prediction model is proposed in [52]. The idea is to predict transient stability based on the power network changes and state variable changes within a short time after faults occur. The STGCN can extract features from these changes and learn the correlation of these changes with power system stability. Only a short-time simulation is required, and the efficiency of stability analysis can be improved.

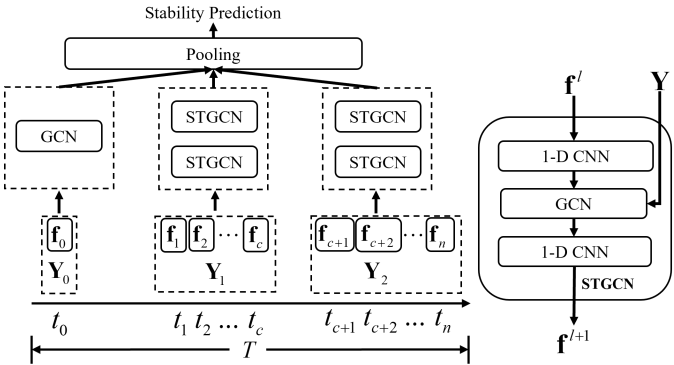

Fig. 4. The architecture of the STGCN-based stability prediction model.

The implementation is supported by the network topology accessibility of Py_PSOPS. The STGCN-based stability prediction model is shown in Fig. 4. Each STGCN layer is composed of a one-dimensional convolutional neural network (1D-CNN) layer, a graph convolutional network (GCN) layer, and another 1D-CNN layer. The input of the $(l+1)$-th layer of STGCN is the feature vector $\mathbf{f}^l$ and the admittance matrix $\mathbf{Y}$, and the output is $\mathbf{f}^{l+1}$. The input features include $\mathbf{Y}_0$, $\mathbf{Y}_1$, and $\mathbf{Y}_2$, i.e., the admittance matrices before the fault, during the fault, and after clearing the fault, respectively, which can be obtained via external solution APIs, and $\mathbf{f}_0, \mathbf{f}_1, ..., \mathbf{f}_n$, i.e., temporal data of selected state variables obtained by a short-time simulation from $t=0$ to $t=T$, which can be obtained using external function APIs. The fault is cleared at the instant $t=t_c$. The total number of sampling instants is $n$. The output of the model is the predicted stability label of the input cases.

After training, the STGCN model and the transient simulation function of the Py_PSOPS can be integrated using APIs and perform efficient transient stability analysis. The pseudo-code of training an STGCN model and integrating the STGCN model with transient simulations is as follows.

| Pseudo-code 2: STGCN-based Stability Prediction |
|---|
| **Training Procedures** |
| **input** the total number of training epochs $N_E$, the training dataset, the test dataset, and the mini-batch size $m$ |
| **initiate** the parameters of the STGCN model |
| **set** $epoch = 1$ |
| **while** $epoch < N_E$ : |
|   **do** |
|     **get** $m$ samples from the training dataset |
|     STGCN.**forward**( $m$ samples) |
|     **calculate** the cross-entropy loss |
|     loss.**backward**() |
|     **update** parameters of STGCN |
|   **until** all the samples in the training dataset have been selected |
|   **evaluate** the STGCN model in the test dataset |
|   **set** $epoch \leftarrow epoch + 1$ |
| **end while** |
| **Integration of the STGCN model and transient simulations** |
| **input** the STGCN model, the contingency, the integration step $\Delta t$, and the short simulation time $T$ |
| **power flow** calculation using **external function APIs** |
| **set** contingency using **internal function APIs** |
| **set** the integration step to $\Delta t$ and the simulation time to $T$ using **internal solution APIs** |
| **transient simulation** using **external function APIs** |
| **get** $Y_0$, $Y_1$, $Y_2$, and $f_0, f_1, ..., f_n$ using **external solution APIs and external function APIs** |
| stability label = STGCN.**forward**( $Y_0$, $Y_1$, $Y_2$, and $f_0, f_1, ..., f_n$ ) |

*2) Test Results*

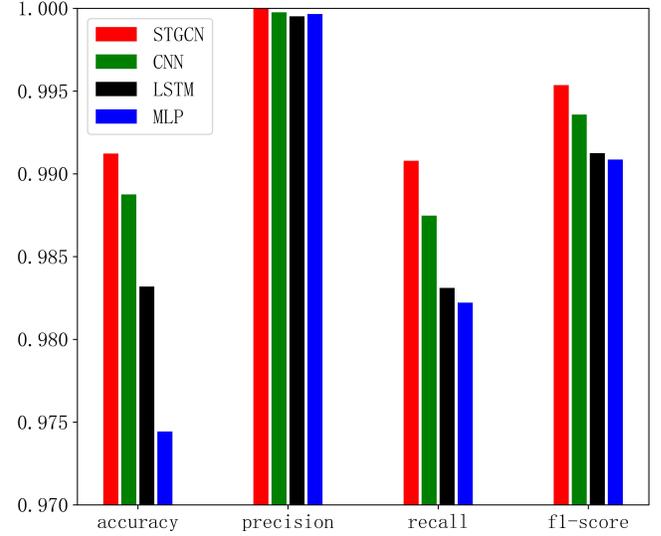

Fig. 5. Results of STGCN、CNN、LSTM、MLP models.

The STGCN-based stability prediction model is trained. Samples in the training dataset are randomly selected in the sample dataset. The training dataset contains 10240 samples, whereas the testing dataset contains 33600 samples. The comparison results of the STGCN model, the convolutional neural network (CNN) model, the long short-term memory (LSTM) model, and the multi-layer perceptron (MLP) model are displayed in Fig. 5, where accuracy, precision, recall, and f1-score are commonly used indices for evaluating the performance of neural networks [52]. Each time of STGCN-based stability prediction averagely costs 5 milliseconds, whereas the complete simulation averagely costs 25 milliseconds.

### C. Neural ODE-based Dynamic Modeling

This is an example of AI-assisted simulation. The trained AI model is integrated into the simulator and is simulated simultaneously with physics-based dynamic models.

*1) Neural ODE-based Power System Dynamic Modeling*

A neural ODE-based dynamic modeling method is proposed in [35]. The idea is to build data-driven dynamic models based on neural ODE and accessible measurement data when there is a lack of prior knowledge of the component, e.g., equivalent modeling of load areas [53] and renewable plants [54]. While learning a global approximation of the derivative functions with easily trainable neural networks, neural ODE also keeps the classical framework of numerical integration, which is a very important prior knowledge and makes neural ODE highly adaptive to scientific computations and industrial applications, as shown in (1):

$$\Psi(\mathbf{x}, \mathbf{V}; \mathbf{\theta}) \doteq \dot{\mathbf{x}} = \mathbf{f}(\mathbf{x}, \mathbf{V}) \quad (1)$$

where $\mathbf{x}$ denotes the state variables, whose time derivatives are equal to $\mathbf{f}(\mathbf{x}, \mathbf{V})$, $\mathbf{V}$ denotes the nodal voltages, $\Psi$ denotes the parametric derivative functions, and $\mathbf{\theta}$ denotes the parameters of the parametric derivative functions. After inputting the initial value $\mathbf{x} = \mathbf{x}(0)$, the variation of $\mathbf{x}$ can be calculated with a numerical integration method. The parameters of neural ODE can be trained using a set of sampled curves of $\mathbf{x}$ and $\mathbf{V}$, which can be obtained through external APIs. The loss function is the mean squared error between the predicted curves and the ground-truth curves of $\mathbf{x}$.

After training, the neural ODE-based dynamic models are directly integrated with physics-based models and transient simulations are carried out to prove the efficacy of the neural model. The pseudo-code of training a neural ODE model and neural ODE model-integrated transient simulations is as follows.

**Pseudo-code 3: Neural ODE-based Power System Dynamic Modeling**

**Training Procedures**
**input** the total number of training epochs $N_E$, as well as the training and test datasets that contain $\{\hat{\mathbf{x}}(t_i), \hat{\mathbf{V}}(t_i)\}, i \in \{0,1,...,n\}$, i.e., the ground-truth values of $\mathbf{x}$ and $\mathbf{V}$ at time instants $t = t_i$
**initiate** the parameters of the neural ODE (NODE) model
**set** $epoch = 1$
**while** $epoch < N_E$:
  **do**
    **get** a sample $\{\hat{\mathbf{x}}(t_i), \hat{\mathbf{V}}(t_i)\}, i \in \{0,1,...,n\}$ from the training dataset
    **set** $\mathbf{x} = \hat{\mathbf{x}}(t_0)$ and $\mathbf{V} = \hat{\mathbf{V}}(t_0)$
    **for** $i$ in $\{1,2,...,n\}$:
      $\mathbf{x}(t_i) = \mathbf{x}(t_{i-1}) + (t_i - t_{i-1}) \cdot \text{NODE}.\mathbf{forward}(\mathbf{x}(t_{i-1}), \hat{\mathbf{V}}(t_{i-1}))$
    **end for**
    **cal** the mean squared loss between $\mathbf{x}$ and $\hat{\mathbf{x}}$
    loss.**backward**()
    **update** the parameters of the NODE model
  **until** all the samples in the training dataset have been selected
  **evaluate** the NODE model in the test dataset
  **set** $epoch \leftarrow epoch + 1$
**end while**

**NODE model-integrated transient simulation**
**input** the NODE model, the contingency, the integration step $\Delta t$, and the simulation time $T$
**set** the component model to the NODE model using **internal model APIs**
**power flow** calculation using **external function APIs**
**set** contingency using **internal function APIs**
**set** the integration step to $\Delta t$ and the simulation time to $T$ using **internal solution APIs**
**transient simulation** using **external function APIs**

*2) Test Results*

The source code of common neural ODE modules for power system dynamic modeling is developed and published on GitHub[2]. A test case of modeling a generator with the published neural ODE module is conducted. The generator at bus 31 is modeled with the classical generator model. There are 3,200 samples in the training dataset and 800 samples in the test dataset. Samples are randomly selected in the sample dataset including stable contingencies and unstable contingencies. The neural model is trained using only the state variables before the maximum rotor angle difference exceeds 360 degrees.

After training, the neural ODE-based dynamic model is integrated into the simulator using the LibTorch Library. The comparative results of the original classical generator model-based simulations and the neural ODE model-integrated simulations are shown in Fig. 6, where $\Delta\delta$ denotes the rotor angle difference between generators at buses 31 and 39, Cal Diff. denotes the difference between the results obtained by the original model and the neural ODE-based model, and abs(Diff.) denotes the absolution value of the difference. As can be seen, the modeling errors are within an acceptable range.

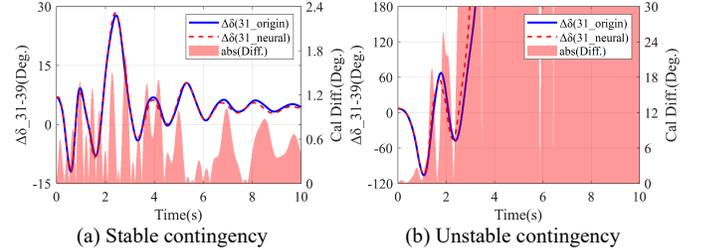

(a) Stable contingency      (b) Unstable contingency

Fig. 6. Comparative results of the classical generator model-based simulations and the neural ODE model-integrated simulations under stable and unstable contingencies.

### D. RL-based SOPF

This is an example of the simulator and AI mutually assisting each other.

*1) Framework Design*

SOPF is one of the traditional control problems of power systems. In SOPF formulation, a target function needs to be optimized under the equality constraints of the power flow equations and the differential-algebraic equations of power system dynamics, as well as the inequality constraints of static security constraints and dynamic security constraints [55], [56]. RL can solve this problem in a simulation-based optimization manner as displayed in Fig. 7. Based on OpenAI Gym [57] and Py_PSOPS, an environment of SOPF solutions is established, which parses actions and outputs state and reward after performing power flow calculation, transient simulation, and constraints check. An AI-based agent determines actions according to the state. The simulator-based environment and the AI-based agent form an interactive mechanism by exchanging states, rewards, and actions.

---

[2] https://github.com/xxh0523/Py_PSNODE.

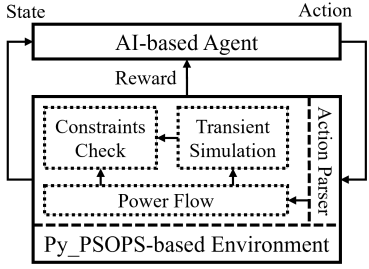

Fig. 7. The framework of the RL-based SOPF.

After training, the AI-based agent can adjust power flow states rapidly and is suitable for online applications. An RL-based optimal power flow solution method has been proposed in [58] using PSOPS and the twin-delayed deep deterministic policy gradient (TD3) algorithm [59]. In this paper, a TD3-based SOPF solution program is realized using Py_PSOPS. The reward design is shown in (2):

$$\text{Reward} = \begin{cases} -999, & \text{if the power flow does not converge} \\ \max[-999, -500 - 50 \times (T_E - T_S)], & \text{if any dynamic constraint is violated} \\ \max[-500, 10 \times L(\mathbf{V}, \mathbf{P}_G)], & \text{if any static constraint is violated} \\ \beta - \alpha C, & \text{if the operating state is secure} \end{cases} \quad (2)$$

where $T_E$ denotes the total simulation time, $T_S$ denotes the time instant when the system loses stability and is set to $T_E$ if the system remains stable, $L(\mathbf{V}, \mathbf{P}_G)$ denotes the sum of the out-of-limit parts of $\mathbf{V}$ and $\mathbf{P}_G$, $C$ denotes the generation cost, and $\alpha$ and $\beta$ represents the scaling factors. The generation cost minimization problem is converted to the maximization problem of the reward. The pseudo-code of RL-based SOPF is as follows.

**Pseudo-code 4: RL-based SOPF**
**input** the total number of training epochs $N_E$ and anticipated contingencies
**initiate** the parameters of the agent
**set** $epoch = 1$
**while** $epoch < N_E$:
   **reset** the operation state using the **Stepwise Power Flow Sampling Scheme**
   **get** the original state using **external function APIs**
   action = Agent.actor.**forward**(state)
   adopt the action and **set** the operation state to the new state using **internal parameter APIs**
   **power flow** calculation using **external function APIs**
   **get** power flow convergence using **external solution APIs**
   **if** power flow calculation converges:
     **get** the new state using **external function APIs**
     **transient simulations** for the anticipated contingencies using **external function APIs**
     **get** the time instants $T_S$ when the system loses stability using **external solution APIs**
     **check** constraints using **external function APIs**
   **end if**
   **calculate** reward
   **update t**he parameters of the agent based on the original state, the actions, the new state, and the reward
   **set** $epoch \leftarrow epoch + 1$
**end while**

*2) Test Results*

The training process is demonstrated in Fig. 8. After the agent is trained, further tests are carried out to check the control effectiveness. 50,000 power flow samples with dynamic constraint violations are sampled. The agent gets the operation state and outputs the control strategy. The agent cost 122.525 seconds, which includes the time consumption of generating a strategy and performing the transient simulation once to check the strategy. After control, 49602 samples return to safe operating points, whereas 398 samples violate static stability constraints. The success rate is 99.204 percent and the new operating points are 100 percent sure to maintain dynamic security.

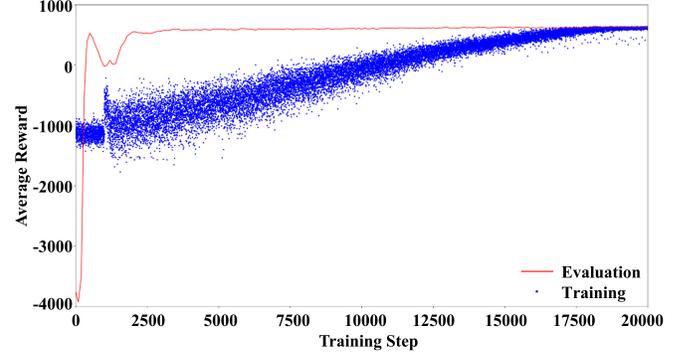

Fig. 8. The training process of the agent with the TD3 algorithm.

## V. CONCLUSION

To conclude, based on the illustration of the interaction mechanism between power system dynamic simulations and AI, an AI-oriented power system transient stability simulator called Py_PSOPS is designed, implemented, tested, and made public. Although it is currently an exploration of AI-oriented power system dynamic simulations, the four test cases demonstrate promising capabilities of Py_PSOPS to support the development of AI-assisted simulations and simulation-assisted AI applications in power system stability analysis and control. It should be noted the development of Py_PSOPS will continue in the future.

**Tannan Xiao** received the B.Eng. and Ph.D. degrees in electrical engineering from Zhejiang University, Hangzhou, China, in 2013 and 2019, respectively. He is currently a Postdoctoral Fellow with the Department of Electrical Engineering, Tsinghua University, Beijing, China. His research interests include power system stability analysis and control, high-performance computing, and artificial intelligence.

**Ying Chen** received the B.S. and Ph.D. degrees in electrical engineering from Tsinghua University, Beijing, China, in 2001 and 2006, respectively, where he is currently a Professor with the Department of Electrical Engineering. His research interests include parallel and distributed computing, electromagnetic transient simulation, cyber-physical system modeling, and cyber security of smart grids.

**Jianquan Wang** received the Ph.D. degree in electrical engineering from Xi'an Jiaotong University, Xi'an, China, in 1997. From 1997 to 1999, he was a postdoctoral fellow with the College of Electrical Engineering, Zhejiang University, Hangzhou, China, where he is currently an Associate Professor. His research interests include power system stability analysis and control, application of artificial intelligence in power systems, and high-performance computing.

**Shaowei Huang** received the B.S. and Ph.D. degrees from Tsinghua University, Beijing, China, in 2006 and 2011, respectively. From 2011 to 2013, he was a Postdoctoral Fellow with the Department of Electrical Engineering, Tsinghua University, where he is currently an Associate Professor. His research interests include power systems modeling and simulation, power system parallel and distributed computing, complex systems and their application in power systems, and artificial intelligence.

**Weilin Tong** received the M.Sc. degree in electrical engineering in 2019 from the College of Electrical Engineering, Zhejiang University, Hangzhou, China. He is currently an Engineer with Wuxi Power Supply Company of State Grid. His research interests include power system hybrid simulation and high-performance computing.

**Tirui He** received his B.Eng. degree in electrical engineering from Tsinghua University, China, in 2021, and the M.Sc. degree in Business Analytics from Nanyang Technology University, Singapore, in 2022. He is currently working as a Data Scientist with Alibaba. His research interests include power system modeling and simulation, and the application of artificial intelligence in power systems.